\documentstyle[multicol,aps,pre,graphicx,amssymb]{revtex}
\draft
\begin{document}
\title{Generalized Quantum Field Theory as an Alternative Approach To The Problem of Composite Particles Reaction }
\author{C.I. Ribeiro-Silva and N. M. Oliveira-Neto \\
Centro Brasileiro de Pesquisas F\'\i sicas,\\
Rua Xavier Sigaud,150 cep 22290-180,\\Rio de Janeiro, Brazil}
\maketitle

\begin{abstract}
A generalization of the Heisenberg algebra has been recently constructed.
 This generalized  algebra has a characteristic function which depends on one of its generators.
 When this function is linear, $qJ_0+s$, it is possible to construct a Generalized Quantum 
Field Theory (GQFT) that creates at a space-time a composite particle. 
In the present work we show that  a generalized QFT can also be constructed  consistently,
 even with a nonlinear characteristic function and leads to better results as long as we 
have more parameters to (possibly) fit the spectrum of a real composite particle.
\end{abstract}


\begin{multicols}{2}

\section{Introduction}
It is well known that the standard quantum field theory (QFT) is constructed 
within the framework of  Heisenberg algebra\cite{heisenberg,moler}. Therefore a possible way to construct
 a non-standard QFT is through generalization of the Heisenberg algebra.
A generalized Heisenberg algebra (GHA) has recently been proposed \cite{paperdeles1}. 
This generalized  algebra has a characteristic function which depends on one of its generators. 
When this function is linear with slope $q$, the algebra turns into a $q$-oscillator \cite{paperdeles1} 
and it was shown \cite{paperdeles2} that, for this case, it is possible to construct a generalized QFT 
that describes at a space-time a composite particle  . In Ref\cite{paperdeles2}  the propagator, 
the first and second order scattering processes were computed and it was shown that 
the convergence of the perturbative series can be changed.
However, other realizations of the GHA have been presented in \cite{paperdeles1,paperdeles3}, 
the case where we have a nonlinear characteristic function. We will show, in this present work, 
that a self-consistent GQFT can also be constructed in that case and  as a consequence  we have  
more parameters to fit experimental data. A detailed discussion comparing these two case will be presented ahead.

The first step is to analyze the GHA for a general characteristic function  $f(J_0)$. 
As defined previously \cite{paperdeles1}, this generalized Heisenberg algebra is  
generated by three operators,  $J_0, A$ and $A^\dagger$, and  described by the following relations

\begin{eqnarray}
J_0 \,  A^\dagger &=& A^\dagger \,  f(J_0) \label{eq:1}\\
A  \, J_0 &=& f(J_0) \,  A \label{eq:2}\\
\left[ A , A^\dagger \right] &=& f(J_0)-J_0  \label{eq:3}
\end{eqnarray}
where $^\dagger$ means Hermitian conjugate and by hypothesis, $ J_{0}^\dagger=J_0 $ 
and $f(J_0)$ is an arbitrary function of $J_0$. Within this algebra we verify that the generators satisfy the Jacobi identity with

\begin{eqnarray}
C = A^\dagger  \,  A - J_0 = A  \, A^\dagger  - f(J_0) \label{eq:4}
\end{eqnarray}
being the Casimir operator of the algebra.
Assuming that there is a vacuum state represented by $|0\rangle$, it can be 
demonstrated \cite{paperdeles1} that for an arbitrary function $f$, that

\begin{eqnarray}
J_0 \, |m-1\rangle &=& f^{(m-1)}(\alpha_0) \, |m-1\rangle,\,\,\, m=1,2 \label{eq:5} \\
A^\dagger \, |{m-1}\rangle &=& N_{m-1} \, |m\rangle \label{eq:6} \\
A \, |{m}\rangle &=& N_{m-1} \, |{m-1}\rangle, \label{eq:7}
\end{eqnarray}
where $N^2_{m-1} = f^m(\alpha_0) - \alpha_0$, $\alpha_0$ is the lowest $J_0$ 
eigenvalue and $f^m(\alpha_0)$ is the $m$th iteration through function $f(\alpha_0)$. 
This GHA describes a class of one-dimensional quantum systems characterized by energy eigenvalues given by

\begin{equation}
\epsilon_n = f(\epsilon_{n-1}) \label{eq:8}
\end{equation}
where $\epsilon_n$ and $\epsilon_{n-1}$ are successive energy levels. 
Unlike standard Heisenberg Algebra (where the energy of the $n$-th level is 
equal to $n$ times the energy of the first one) ,here, the energy of the 
$n$-th level (depending on the value of the parameters) is greater or smaller 
than $n$ times the energy of the first one. In the last case, the energy gap 
between consecutive levels becomes smaller as n increases, behaving like the energy spectrum of a composite particle.
So we can postulate that this algebra describes a composite particle.

\section{A Generalized QFT}

Let us discuss, firstly, the algebra given by  (\ref{eq:1})-(\ref{eq:3}) for the  
quadratic case, i.e. $f(J_0) = t \, J^2_0 + q \, J_0 + s$, which is the simplest 
nonlinear case. The algebraic relations (\ref{eq:1})-(\ref{eq:3})  can be written as \cite{paperdeles1}

\begin{eqnarray}
\left[ J_0 , A^\dagger  \right]_{q} &=& t  \, A^\dagger J^2_0 + s \, A^\dagger,  \label{eq:9} \\
\left[ J_0 , A \right]_{q^{-1}} &=& -\frac{t}{q} \, J^2_0 \, A - \frac{s}{q} \, A,\\
\label{eq:10}
\left[ A^\dagger , A  \right] &=& t  \, J^2_0 + ({q}-1) \, J_0 + s,  \label{eq:11}
\end{eqnarray}
where $\left[a,b \right]_q= a \, b-q \, b \, a$, is the $q$-deformed commutation relation of  two operators $a$ and $b$.
The relations  (\ref{eq:9})-(\ref{eq:11}) describe  a two-parameter deformed Heisenberg Algebra already studied\cite{paperdeles1}.
Of course, for $t=0$ we recover the linear case  and if additionely $q=1$, the standard Heisenberg algebra.

We focus now on  the graphical analysis of the function $f(\alpha_0)=t\alpha_0^2+q\alpha_0+s$. 
Let us plot $y=f(\alpha_0)$ together with $y=\alpha_0$. In the points where lines 
intersect, we have $\alpha_0=y=f(\alpha_0)$. So the intersections are precisely the fixed points.
\begin{figure}
\begin{center}
\includegraphics[angle=0.0,scale=0.3]{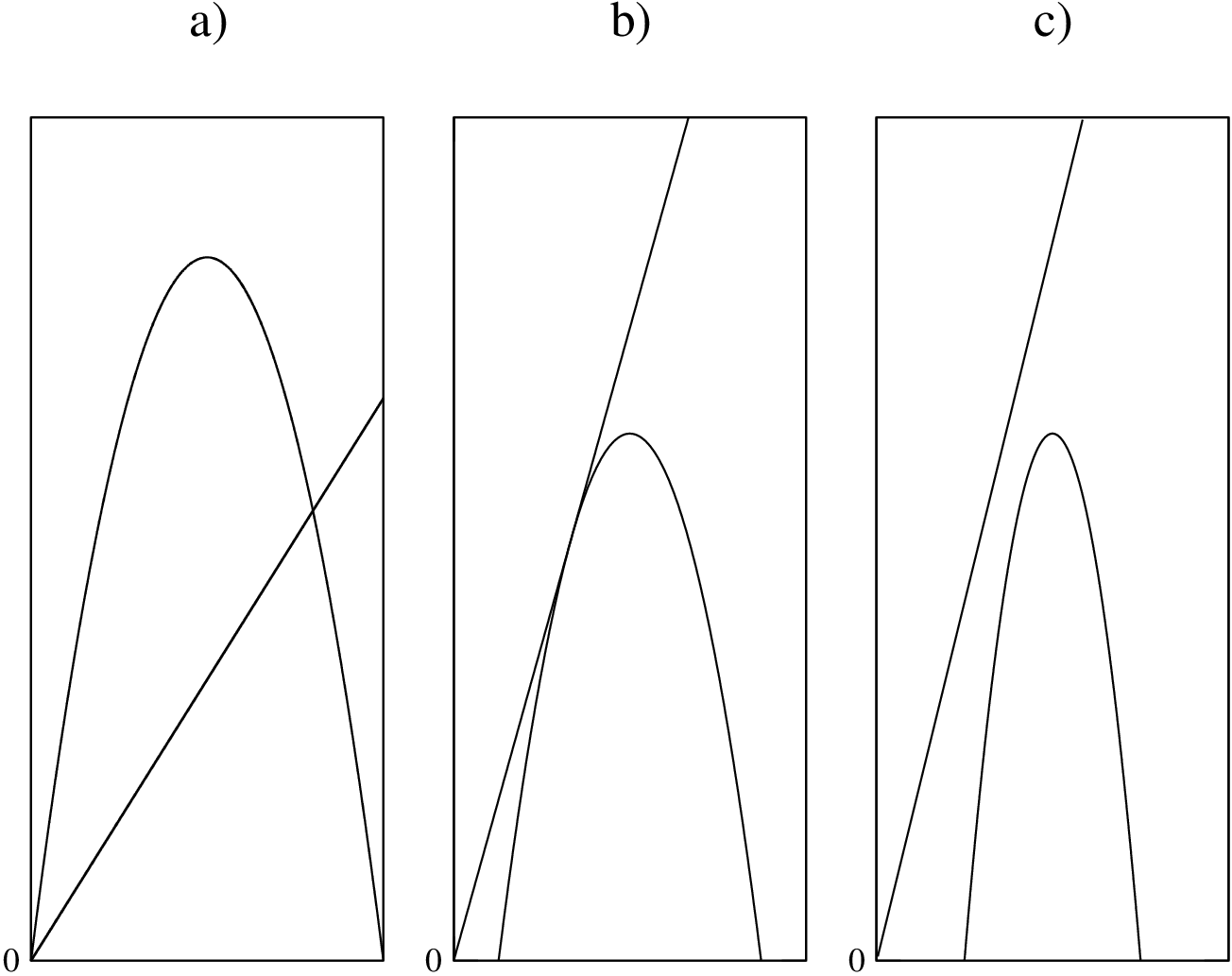}
\caption{ (a) $\Delta>0$ (b) $\Delta=0$ and (c) $\Delta<0$, where $\Delta=(q-1)^2+4ts$. }
\end{center}
\end{figure}
Assuming $t<0$, there are three cases to be analyzed: $ (a)$  $\Delta 
< 0$ $(b)$ $\Delta=0$ and $(c)$ $\Delta > 0$, for $\Delta = (q-1)^2+4ts$ 
(see Fig 1). However, as depicted in (Fig. 1), depending on $t,q,s$ and 
$\alpha_0$, we have different spectra. Nevertheless, we are interested 
in a special spectrum which can be associated to a composite particle, 
thus, only the case $(c)$ for $\alpha_{min}<\alpha_0<\alpha_{max}$ is relevant.
Fig 2 shows the comparison between the quadratic and the linear cases. 

As one may notice, in the linear case, there is a simple relation among
energy level gaps ( as $n$ increases, the energy gap between sucessive levels  always decreases) wich make the linear case unsuitable to fit some realistic  spectra, see for instance\cite{nucleoxe}. This makes the quadratic more suitable to fit the 
spectra of the mainstream  composite particle.
\vspace{5cm}
\begin{figure}\nonumber
\begin{center}
\includegraphics[angle=0.0,scale=0.3]{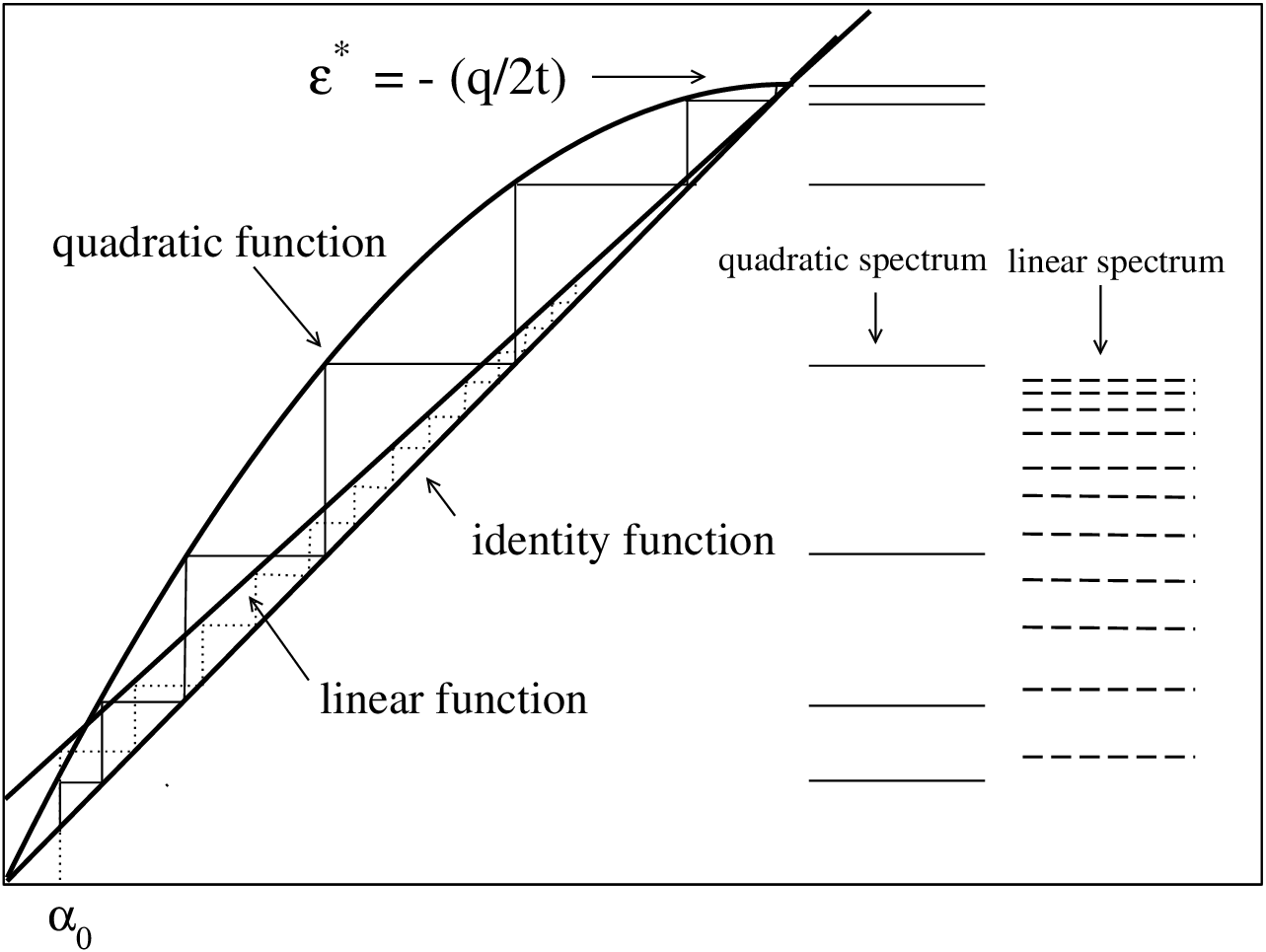}
\caption{Comparison between the quadratic and  linear cases}
\end{center}
\end{figure}
\vspace{1cm}
Now, we will realize the operators $A^\dagger$, $A$ and $J_0$ in terms of 
physical operators, as in the case of the one-dimensional harmonic oscillator
 \footnote{where $A^\dagger$ and $A$ can  be written as a function of $p$ and $q$.}.
 In order to do so we will employ the formalism of non-commutative differential 
and integral calculus \cite{dimakis}  which considers a one-dimensional lattice in a 
momentum space where the momenta are allowed to take only discrete values $p_0$, $p_0+a$, $p_0+2a$, 
and so on, with $a > 0$. Let  us introduce the momentum shift operator

\begin{eqnarray}
T = 1 + a \, \partial_p \label{eq:12} \\
\bar{T} = 1 - a \, \bar{\partial}_p \label{eq:13}
\end{eqnarray}
where ${\partial}_p$ and $\bar{\partial}_p$ are the left and right discrete derivatives that shift the momentum value by $a$, i.e.

\begin{eqnarray}
T \, f(p) = f(p+a) \label{eq:14}, \\
\bar{T} \, f(p) = f(p-a) \label{eq:15}
\end{eqnarray}
and satisfy

\begin{equation}
T \, \bar{T} = \bar{T} \, T = \hat{1} \label{eq:16}. 
\end{equation}
Introducing the momentum operator $P$

\begin{equation}
P \, f(p) = p \,  f(p) \label{eq:17},
\end{equation}
hence

\begin{eqnarray}
T \, P = (P+a) \, T \label{eq:18} \\
\bar{T} \, P = (P-a) \, \bar{T} \label{eq:19}.
\end{eqnarray}

Now, we return to the realization, observing that, in this case we have not 
an explicit formula for $f^m(\alpha_0)$ as in the linear one \cite{paperdeles1}, 
but we can still associate this two-parameter deformed Heisenberg algebra (\ref{eq:9})-(\ref{eq:11}) 
to the one-dimensional lattice we have just presented. By defining an operator $N$ such that

\begin{equation}
N|m\rangle = m|m\rangle, 
\end{equation}
we can write an operator $f(N,\alpha_0)$ that verifyes 

\begin{equation}
f(N,\alpha_0)|m\rangle = f^m(\alpha_0)|m\rangle,
\end{equation}
So, using (\ref{eq:5})-(\ref{eq:7}) we can write $J_0$ as  

\begin{equation}
J_0 \equiv f(N,\alpha_0) =  \alpha_N \equiv f({P/a},\alpha_0)  \label{eq:21} \\
\end{equation}
where we heve defined $N = P/a$, with P given by (\ref{eq:17}). This implies that

\begin{equation}
 P \, |m\rangle =  m \, a \, |m\rangle,\,\,\,\,\, m=0,1,2,... \label{eq:23} \\
\end{equation}
Moreover, it is easy to see \cite{paperdeles2} that 

\begin{equation}
 \bar{T} \, |m\rangle =  |{m+1}\rangle,\,\,\,\,\, m=0,1,2,... \label{eq:24} \\
\end{equation}
with $\bar{T}$ and $T = \bar{T}^\dagger$  defined in Eqs.(\ref{eq:14})-(\ref{eq:15}).

Now, using Eqs.(\ref{eq:5})-(\ref{eq:7}), we finally define

\begin{eqnarray}
A^\dagger & =& S(P) \, \bar{T}, \label{eq:25} \\
A &=& T \, S(P), \label{eq:26}
\end{eqnarray}
where
\begin{equation}
S(P)^2 = J_0 - \alpha_0, \label{eq:27}
\end{equation}
being  $\alpha_0$  the lowest $J_0$ eigenvalue.
Following similar steps to those used to construct a standard spin-$0$ QFT \cite{livro}, let us define  two  operators 
{\small
\begin{eqnarray}
\chi &\equiv& \imath \, (S(P) \, (1-a \, \bar{\partial}_p)-(1+a \, \partial_p) \, S(P))=-\imath  (A-A^\dagger) \label{eq:28} \\
Q &\equiv& S(P) \, (1-a \, \bar{\partial}_p)+(1+a \, \partial_p) \, S(P) =A+A^\dagger , \label{eq:29}
\end{eqnarray}
}
satisfying

\begin{eqnarray}
\left[ \chi , P \right] &=&-\imath \, a \, Q    \label{eq:30} \\
\left[ P , Q \right] &=& \imath \, a \, \chi   \label{eq:31} \\
\left[ \chi , Q \right] &=& 2 \, \imath (S^2(P)-S^2(P+a)).  \label{eq:32}
\end{eqnarray}

At this point, we introduce an independent copy of the one-dimensional momentum lattice,
 we have just defined, at each point of  a  $\vec K$-lattice through the substitution $P\rightarrow P_{\vec k}$ so,

\begin{eqnarray}
A^\dagger_{\vec{k}} & =& S_{\vec{k}} \, \bar{T}_{\vec{k}}, \\
A_{\vec{k}}&=& T_{\vec{k}} \, S_{\vec{k}}, \\
J_0(\vec{k})&=&f^\frac{P_{\vec{k}}}{a}(\alpha_0).
\end{eqnarray}

Then, we define three field operators 

\begin{eqnarray}
\phi(\vec{r},t) &=& \sum^{}_{\vec{k}} \frac{1}{\sqrt{2 \, \Omega \, w(\vec{k})} } 
\, (A_{\vec{k}}^{\dagger} \, e^{-\imath \, \vec{k}. \, \vec{r}} + A_{\vec{k}} \, e^{\imath \, \vec{k}. \, \vec{r}}) ,   \label{eq:33} \\
\Pi(\vec{r},t) &=& \sum^{}_{\vec{k}} \frac{\imath \, w(\vec{k})}{\sqrt{2 \, \Omega \, w(\vec{k})} } \, (A_{\vec{k}}^{\dagger} 
\, e^{-\imath \, \vec{k}. \, r} - A_{\vec{k}} \, e^{\imath \, \vec{k}. \, \vec{r}}) ,  \label{eq:34}\\ 
\wp(\vec{r},t) &=& \sum^{}_{\vec{k}} \sqrt{ \frac{w(\vec{k})}{2 \, \Omega} } \,  S_{\vec{k}} \, e^{-\imath \, \vec{k}. \, \vec{r}},\label{eq:35}  
\end{eqnarray}
where $w(\vec{k})=\sqrt{\vec{k}^2 + m^2}$ and $m$ is a real number.
Using (\ref{eq:33})-(\ref{eq:35}) we can show that the Hamiltonian
{\small
\begin{eqnarray}
H = \frac{1}{2} \, \int d^3 \,\vec{r} (\Pi^2(\vec{r},t) + u |\wp(\vec{r},t)|^2 \\ \nonumber
+\phi(\vec{r},t) \, (-\vec{\nabla}^2+ m^2) \, \phi(\vec{r},t) ) 
\end{eqnarray} 
}
can be written as

\begin{equation}
H = \frac{1}{2} \, \sum^{}_{\vec{k}} w(\vec{k}) \, \left[S^2_{\vec{k}}(N+1)+ (1+u) S^2_{\vec{k}}(N)\right] \label{eq:37}
\end{equation}
where $N_0^2 = t{\alpha_0}^2+(q-1)\alpha_0+1$ and $S^2(N)$ is given by  Eq.(\ref{eq:27}). 
Note that in the limit $t\rightarrow 0$, we recover the linear case and  for 
 $t\rightarrow 0$, $q \rightarrow 1 (u \rightarrow 0)$, the Hamiltonian is proportional to the number operator.

The time evolution of the fields can be studied by solving Heisenberg's equation for $A_{\vec{k}}^{\dagger},A_{\vec{k}}  $.
So, using Eq. (\ref{eq:37}) we have

\begin{equation}
\left[ H,A_{\vec{k}}^{\dagger} \right] = w(\vec{k}) \, A_{\vec{k}}^{\dagger} \, h(N_{\vec{k}}) \label{eq:38}
\end{equation}
where for the quadratic case 

\begin{equation}
 h(N_{\vec{k}})= \frac{1}{2}\Delta E\left[t(S^2_{\vec{k}}(N+1)+ S^2_{\vec{k}}(N))+2t\alpha_0+Q\right] ,\label{eq:39}
\end{equation}
with $Q=1+u+q$ and $\Delta E =S^2_{\vec{k}}(N+1)- S^2_{\vec{k}}(N)$ . For the general case $f(\alpha_0) = \sum^{n}_{j=0} a_j\alpha_0^j$, we have 
{\small
\begin{equation}
h(N_{\vec{k}}) = \frac{1}{2}\Delta E\sum^{n}_{j=1}\sum^{j-1}_{i=0}\left[a_j(S^2_{\vec{k}}(N)+\alpha_0)^{j-1}\chi(N_{\vec{k}})^i\right]+(u+1)
\end{equation}
}
with 

\begin{equation}
\chi(N) = \frac{S^2_{\vec{k}}(N+1)+\alpha_0}{S^2_{\vec{k}}(N)+\alpha_0}.
\end{equation}

Solving the Heisenberg equation 

\begin{equation}
A_{\vec{k}}^{\dagger}  = A_{\vec{k}}^{\dagger}(0) e^{\imath \, w(\vec{k}) \, h(N_{\vec{k}}) \, t} \label{eq:40}
\end{equation}
we  can write Eq.(\ref{eq:33}) as $\phi(\vec{r},t) = \alpha(\vec{r},t) + \alpha^\dagger(\vec{r},t)$, 
 \footnote{hereafter $A_{\vec{k}}^{\dagger}(t=0) \equiv A_{\vec{k}}^{\dagger}$}
where
\begin{equation}
\alpha^\dagger(\vec{r},t) = \sum^{}_{\vec{k}} \frac{1}{\sqrt{2 \, \Omega \, w(\vec{k})} } 
\, A_{\vec{k}}^{\dagger} \,  e^{-\imath \, \vec{k}.\vec{r}+ \imath \, w(\vec{k}) \, h(N_{\vec{k}}) \, t} .   \label{eq:41}
\end{equation}
 
The Feynman propagator $D^{N}_{F}(x_1,x_2)$ defined as Dyson-Wick contraction 
between ($x_i \equiv (r_i,t_i)$) $\phi(x_1)$ and $\phi(x_2)$ can be 
computed using  (\ref{eq:41}). In the integral representation it is given by
{\small
\begin{eqnarray}
D^{N}_{F}(x) &=& \frac{-\imath}{(2 \, \pi)^4} \, \int d^4 \,k \frac{S^2_k(N+1) \, e^{\imath \, \vec{k}.\vec{r}- \imath \, k_0 \, h(N_{\vec{k}})} \, t }{k^2 + m^2}+\\ \nonumber 
&+& (N \rightarrow N-1).
\end{eqnarray}
}
Note that, if $q \rightarrow 1$ the standard result is recovered.

\section{Perturbative Computation}

We shall now analyze the scattering process 
$1+2 \rightarrow 1^{'}+2^{'}$ for $\vec{p}_1 \neq \vec{p}_2 \neq \vec{p^{'}}_1 \neq \vec{p^{'}}_2$ with initial state

\begin{equation}
N^2_0 \, |1,2\rangle \equiv A^{\dagger}_{\vec{p_1}} \,  A^{\dagger}_{\vec{p_2}} \, |0\rangle\label{eq:43}
\end{equation}
and final state
\begin{equation}
N^2_0 \, |1,2\rangle \equiv A^{\dagger}_{{\vec{p_1}}^{'}} \,
  A^{\dagger}_{{\vec{p_2}}^{'}} \, |0\rangle\label{eq:44}
\end{equation}
where these particles are described by the Hamiltonian given 
in Eq.(\ref{eq:37}) with an interaction $\lambda \, \int :\phi^4(\vec{r},t): d^3 \, \vec{r}$. 

For the first order scattering process we have 
{\small
\begin{equation}
 {S_{fi}}_{(1)} = \frac{-6 \, (2 \,\pi)^4 \, \imath \, N^{2}_0  \, \lambda \, \delta^4(P_{1}+P_{2}-P'_{1}-P'_{2})  }
{\Omega^2 \, (t \, (N^{2}_0 + 2 \, \alpha_0) + Q) \, 
\sqrt{\omega_{\vec{p}_1}\omega_{\vec{p}_2}\omega_{\vec{p'}_1}\omega_{\vec{p'}_2} } 
}  \label{eq:45}
\end{equation}
}
and for the second order 
{\small
\begin{equation}
 {S_{fi}}^{a}_{(2)}=  
\frac{ \, N^4_0 \, \lambda^2 \,\, \delta^4(P_{1}+P_{2}-P'_{1}-P'_{2}) \, I }
{2 \,  \Omega^2 \, (t \, (N^{2}_0 + 2 \, \alpha_0) + Q)^2 \, 
\sqrt{\omega_{\vec{p}_1}\omega_{\vec{p}_2}\omega_{\vec{p'}_1}\omega_{\vec{p'}_2} } 
} \label{eq:46}
\end{equation}

\begin{equation}
 {S_{fi}}^{b}_{(2)}= \frac{ \, N^4_0 \, \lambda^2 \,\, \delta^4(P_{1}+P_{2}-P'_{1}-P'_{2}) \, I^\prime }
{2 \,  \Omega^2 \, (t \, (N^{2}_0 + 2 \, \alpha_0) + Q)^2 \, 
\sqrt{\omega_{\vec{p}_1}\omega_{\vec{p}_2}\omega_{\vec{p'}_1}\omega_{\vec{p'}_2} } 
} \label{eq:47}
\end{equation}

\begin{equation}
 {S_{fi}}^{c,t}_{(2)} = \frac{ \, N^4_0 \, \lambda^2 \,\, \delta^4(P_{1}+P_{2}-P'_{1}-P'_{2}) \, I^{\prime\prime} }
{8 \,  \Omega^2 \, (t \, (N^{2}_0 + 2 \, \alpha_0) + Q)^2 \, 
\sqrt{\omega_{\vec{p}_1}\omega_{\vec{p}_2}\omega_{\vec{p'}_1}\omega_{\vec{p'}_2} } 
} \label{eq:48}
\end{equation}

\begin{equation}
 {S_{fi}}^{c,u}_{(2)}= \frac{ \, N^4_0 \, \lambda^2 \,\, \delta^4(P_{1}+P_{2}-P'_{1}-P'_{2}) \, I^{\prime\prime\prime} }
{8 \,  \Omega^2 \, (t \, (N^{2}_0 + 2 \, \alpha_0) + Q)^2 \, 
\sqrt{\omega_{\vec{p}_1}\omega_{\vec{p}_2}\omega_{\vec{p'}_1}\omega_{\vec{p'}_2} } 
} \label{eq:49}
\end{equation}
}
where,

\begin{eqnarray}
P_{i} &=& (\vec{p}_i,\omega_{\vec{p}_i}), \\
P'_{i} &=& (\vec{p'}_i,\omega_{\vec{p'}_i}), \\
I&=& \int d^4 \, k \, \frac{1}{(k^2 + m^2) \, ((-k+s)^2 + m^2)}\, ,\\
\end{eqnarray}
with $s=P_{1}+P_{2}$, and 

\begin{eqnarray}
I^{'}&=& I(s \rightarrow - \, s ), \\
I^{''}&=& I(s \rightarrow  t ), \\
I^{'''}&=& I(s \rightarrow  u ), 
\end{eqnarray}
being

\begin{eqnarray}
t&=&P_{1}-P_{1'} ,\\
u&=&P_{1}-P_{2'}.
\end{eqnarray}

So, up to second order we have
{\small
\begin{eqnarray}
S_{fi}&=& \frac{\lambda \, N^{2}_0}{(N^{2}_0 + 2 \, \alpha_0) + Q} \, A_1 +\\ \nonumber 
&+&\frac{\lambda^{2} \, N^{4}_0} {((N^{2}_0 + 2 \, \alpha_0) + Q)^2} \, (A^{s}_2 +A^{t}_2 +A^{u}_2 )
\end{eqnarray} 
}
where $A_1, A^{s}_2, A^{t}_2$ and $A^{u}_2$  are the same contributions 
that one can find in the structureless particle  standard $\lambda$-$ \phi^{4}$ 
theory model corresponding to the $s,t$ and $u$ channels for one-loop level respectively.
\begin{figure}\nonumber
\begin{center}
\includegraphics[angle=0.0,scale=0.3]{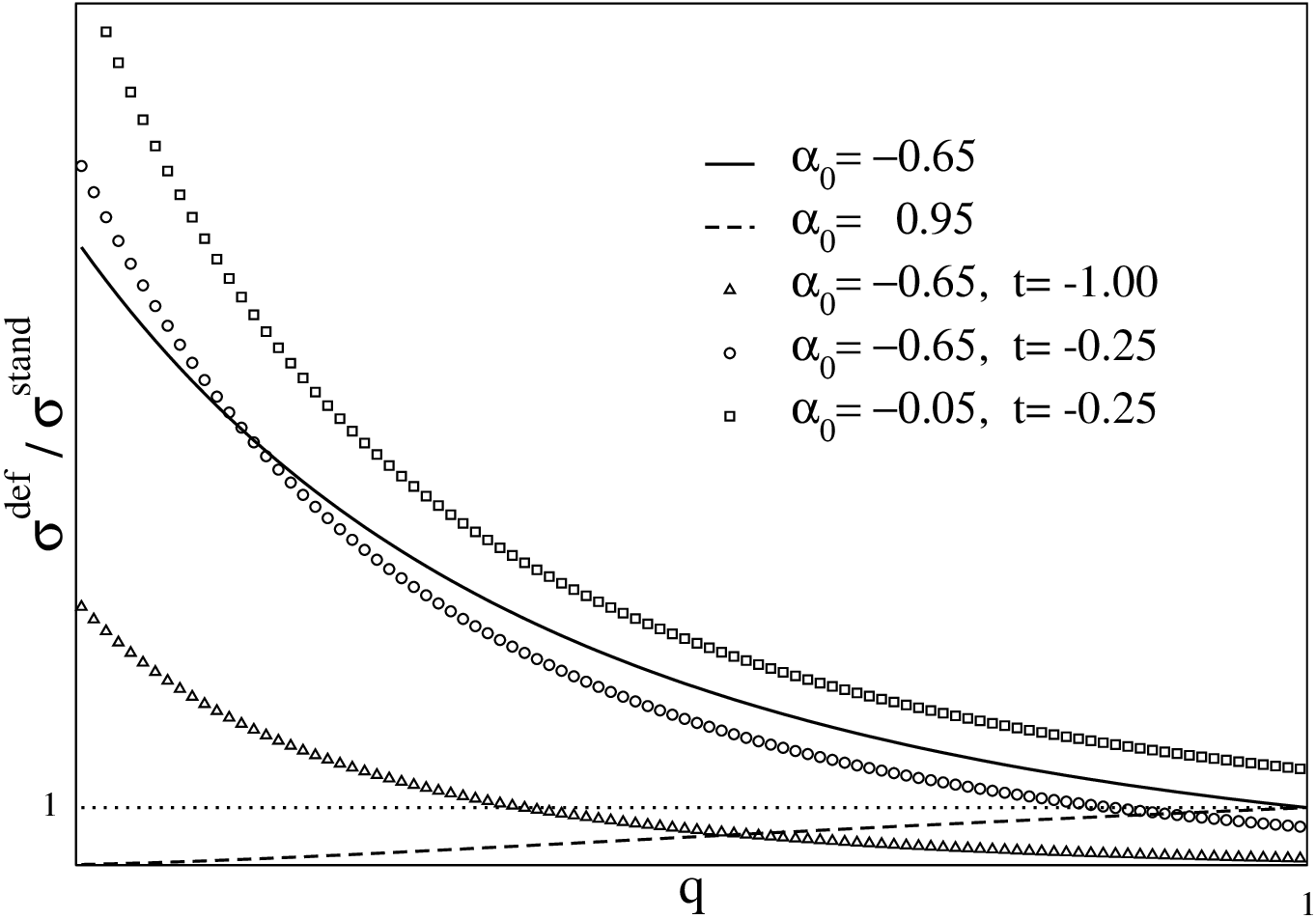}
\caption{Comparison between the deformed cross sections 
for quadratic (symbols) and linear (lines) cases with the standard one. }
\end{center}
\end{figure}
Fig. 3 compares the deformed cross section, for the linear and quadratic cases,
 with the standard one\cite{greiner}. As one can see, the linear case always gives 
a deformed-cross section smaller than the standard one  for $0< \alpha_0 < \epsilon^{*}$
 and greater for $\alpha_0<0$ (for $0<q<1$). Therefore, it becomes difficult to fit both 
energy espectrum of the composite particle and cross section, inasmuch as we may have an 
ambiguous situation where we need  $\alpha_0<0$ to fit the energy spectrum and  $\alpha_0>0$ 
for the cross section. However, in the quadratic case  we do not have such 
ambiguity (as depicted in Fig. 3) because we have one more parameter. 

\section{Conclusion}
We showed that within the framework of deformed Heisenberg algebra, 
with a quadratic characteristic function,  it is possible to construct a generalized 
quantum field theory (GQFT) that describes a composite particle. Comparison between 
a GQFT made with a linear\cite{paperdeles2} and quadratic characteristic function  
was performed showing that, the latter, is more suitable to fit 
experimental data. It is worthwhile to mention that a GQFT made with a  general 
characteristic function brings wider possibilities than the quadratic one but 
several restrictions  must be imposed to the parameters, $ a_1,a_2,...,a_N $, 
in order to describe a composite particle. These restrictions are already present
 in the quadratic case but the algebra is much more simple for this case.  
One can show that the general case can be obtained replacing  
$Q \rightarrow \sum^{n}_{j=1}\sum^{j-1}_{i=0}\left[a_j(\alpha_0)^{j-1}( \frac{N_0^2+\alpha_0}{\alpha_0})^i\right]+(u+1)$. 
In future works we will address this formalism to analize  Compton scattering by 
nuclei below pion threshold where no quantitative consistent description  exists based on first principles\cite{nuclear}.   

\vspace{1cm}
\noindent
{\bf Acknowledgments:} The authors thank  CNPq (Brazil) 
for partial support, Evaldo Curado and Marco Aurelio Rego-Monteiro for valuable comments

\end{multicols}

\end{document}